\documentclass[10pt, twocolumn,secnumarabic,amssymb,amsmath,nobibnotes,aps,prl]{revtex4}
\usepackage{epsf,epsfig,latexsym}
\usepackage{amsmath}
\usepackage{amssymb}
\usepackage{bm}
\usepackage{graphicx}
\usepackage{dcolumn}
\usepackage{multirow}
\usepackage{units}

\expandafter\ifx\csname package@font\endcsname\relax\else
 \expandafter\expandafter
 \expandafter\usepackage
 \expandafter\expandafter
 \expandafter{\csname package@font\endcsname}%
\fi

\def\ssNN#1{\sqrt{s_{NN}} \ifx|#1|\else=\unit[#1]{GeV}\fi}
\DeclareRobustCommand{\snn}[1]{\ifmmode\ssNN{#1}\else$\ssNN{#1}$\fi}

\begin{document}

\title{Rapidity dependence of high $p_T$ suppression at
  $\sqrt{s_{NN}}=\unit[62.4]{GeV}$.}

\newcommand{\bnl}           {$\rm^{1}$}
\newcommand{\ires}          {$\rm^{2}$}
\newcommand{\kraknuc}       {$\rm^{3}$}
\newcommand{\krakow}        {$\rm^{4}$}
\newcommand{\baltimore}     {$\rm^{5}$}
\newcommand{\newyork}       {$\rm^{6}$}
\newcommand{\nbi}           {$\rm^{7}$}
\newcommand{\texas}         {$\rm^{8}$}
\newcommand{\bergen}        {$\rm^{9}$}
\newcommand{\bucharest}     {$\rm^{10}$}
\newcommand{\kansas}        {$\rm^{11}$}
\newcommand{\oslo}          {$\rm^{12}$}

\author{
  I. Arsene\bucharest, 
  I.~G.~Bearden\nbi, 
  D.~Beavis\bnl, 
  C.~Besliu\bucharest, 
  B.~Budick\newyork, 
  H.~B{\o}ggild\nbi, 
  C.~Chasman\bnl, 
  C.~H.~Christensen\nbi, 
  P.~Christiansen\nbi, 
  J.~Cibor\kraknuc, 
  R.~Debbe\bnl, 
  E.~Enger\oslo,  
  J.~J.~Gaardh{\o}je\nbi, 
  K.~Hagel\texas, 
  H.~Ito\bnl, 
  A.~Jipa\bucharest, 
  F.~Jundt\ires, 
  E.B.~Johnson\kansas,
  J.~I.~J{\o}rdre\bergen, 
  C.~E.~J{\o}rgensen\nbi, 
  R.~Karabowicz\krakow, 
  E.~J.~Kim{$\rm^{1,11}$}, 
  T.~Kozik\krakow, 
  T.~M.~Larsen\nbi, 
  J.~H.~Lee\bnl, 
  Y.~K.~Lee\baltimore, 
  S.~Lindal\oslo, 
  R.~Lystad\bergen,
  G.~L{\o}vh{\o}iden\oslo, 
  Z.~Majka\krakow, 
  M.~Mikelsen\oslo, 
  M.~Murray{$\rm^{8,11}$}, %
  J.~Natowitz\texas, 
  B.~Neumann\kansas, 
  B.~S.~Nielsen\nbi, 
  D.~Ouerdane\nbi, 
  R.~P\l aneta\krakow, 
  F.~Rami\ires, 
  C.~Ristea\nbi, 
  O.~Ristea\bucharest, 
  D.~R{\"o}hrich\bergen, 
  B.~H.~Samset\oslo, 
  D.~Sandberg\nbi, 
  S.~J.~Sanders\kansas, 
  R.~A.~Scheetz\bnl, 
  P.~Staszel\krakow, 
  T.~S.~Tveter\oslo, 
  F.~Videb{\ae}k\bnl, 
  R.~Wada\texas, 
  Z.~Yin\bergen, 
  H.~Yang\bergen, and
  I.~S.~Zgura\bucharest \\ 
  The BRAHMS Collaboration \\ [1ex]
  \bnl~Brookhaven National Laboratory, Upton, New York 11973 \\
  \ires~Institut de Recherches Subatomiques and Universit{\'e} Louis
  Pasteur, Strasbourg, France\\
  \kraknuc~Institute of Nuclear Physics, Krakow, Poland\\
  \krakow~Smoluchkowski Inst. of Physics, Jagiellonian University, Krakow, Poland\\
  \baltimore~Johns Hopkins University, Baltimore 21218 \\
  \newyork~New York University, New York 10003 \\
  \nbi~Niels Bohr Institute, Blegdamsvej 17, University of Copenhagen, Copenhagen 2100, Denmark\\
  \texas~Texas A$\&$M University, College Station, Texas, 17843 \\
  \bergen~University of Bergen, Department of Physics, Bergen, Norway\\
  \bucharest~University of Bucharest, Romania\\
  \kansas~University of Kansas, Lawrence, Kansas 66045 \\
  \oslo~University of Oslo, Department of Physics, Oslo, Norway\\
 }

\begin{abstract}
We present measurements of charged hadron $p_T$ spectra from
$Au+Au$ collisions at \snn{62.4} for pseudorapidities $\eta=0$,
$\eta=1$ and $\eta=3.2$. Around midrapidity ($\eta=0$, $\eta=1$)
we find nuclear modification factors at levels suggesting a
smaller degree of high $p_T$ suppression than in the same reaction
at higher energy \snn{200}. At the high pseudorapidity,
$\eta=3.2$, where nuclear modification factors cannot be
constructed due to the lack of $p+p$ reference data, we find a
significant reduction of $R_{CP}$ (central to peripheral ratio) as
compared to midrapidity.  \\
\end{abstract}

\maketitle


The observation of the suppression of the high transverse momentum
part ($p_T > \unit[2]{GeV/c}$) of particle spectra from central
collisions between gold ions relative to similar (scaled) spectra from
proton-proton collisions has been one of the central discoveries at
the Relativistic Heavy Ion Collider, RHIC~\cite{whitepapers}.

For particles emitted around midrapidity large suppression factors
(of order 3-5) have been observed for charged and neutral hadrons
in central $Au+Au$ collisions at
\snn{200}~\cite{brahms_highpt,star_rauau_200,phenix_rauau_200,phobos_rauau_200}.
Similar studies comparing spectra from $d+Au$ collisions to the
same reference $p+p$ collisions, on the contrary, show a lack of
suppression~\cite{brahms_highpt,phenix_dau_highpt,
phobos_dau_highpt,star_dau_highpt}. In fact, in the $d+Au$
collisions, an enhancement of the particle yield in the
$\unit[2]{GeV/c}\lesssim p_T \lesssim\unit[8]{GeV/c}$ range is
found. Current theory explains the observed phenomena in terms of
scattering and interaction between quarks and gluons in the
moments between first contact of the colliding nuclei and jet
fragmentation (i.e.\ formation of hadrons) in subsequent stages of
the collision. The high $p_T$ suppression observed in central
nucleus--nucleus collisions is thought primarily to be the result
of the energy loss of hard scattered partons as they propagate
through a medium with a high density of unscreened color charges.
The theory of the strong interaction, Quantum Chromo Dynamics
(QCD), predicts a high degree of energy loss of scattered partons
due to stimulated gluon emission (in a manner proportional to the
square of the distance
traversed)~\cite{jet_quench1,jet_quench2,jet_quench3}. Conversely,
the enhancement seen in $d+Au$ collisions, in which an extended
dense absorbing medium is believed \emph{not} to be produced, is
thought to be due to multiple scattering between partons (leading
to a broadening of measured $p_T$ distributions), the so--called
Cronin effect.

For particles emitted at forward rapidity ($\eta > 2$) large
suppression factors have also been observed at \snn{200} in
$Au+Au$ collisions~\cite{brahms_highpt,hardprobes2004} as well as
in $d+Au$ collisions at \snn{200}~\cite{brahms_rda_vs_eta}. The
underlying mechanism is at present unclear. An intriguing
possibility relies on the fact that particles observed at large
rapidities, i.e.\ small angles relative to the beam direction, may
originate from scatterings involving a gluon carrying a small
fraction of the nucleon momentum (low--x gluon). The theory of the
Color Glass Condensate~\cite{CGCtheory} predicts that the number
of such low--x gluons may be limited due to fusion among highly
delocalized gluons. This, in turn, has been predicted to lead to
an overall reduction of the yield of charged particle seen at
forward rapidity in $d+Au$ collisions on the $Au$ 'shadow side',
i.e. for collisions predominantly involving a low--x gluon from
the gold and a higher momentum parton from the deuteron.

In the present letter we extend these studies to $Au+Au$
collisions at \snn{62.4} and investigate to what extent the
suppression phenomena observed at the top energy of RHIC persist
at lower CM energies. The PHOBOS experiment has presented studies
of high $p_T$ production for $Au+Au$ at \snn{62.4} averaged over
an interval of pseudorapidity $\eta=0.2-0.4$~\cite{phobos_62}. The
present measurement addresses the pseudorapidity dependence of the
high $p_T$ particle production in $Au+Au$ collisions at this
energy with measurements in narrow intervals ($\Delta \eta = 0.2$)
around $\eta=0,1$ and $3.2$.

Measurements of high $p_T$ particle production at lower energy
(i.e. at the SPS, \snn{17.2}) are sparse and the interpretation of
the data is at present unclear. Depending on the chosen
nucleon--nucleon spectrum to compare to, the nuclear modification
factors obtained in the SPS energy range vary from a large
enhancement~\cite{wang_sps_rhic_study,wa98_raa,ceres_raa} in the
range $p_T$ = 2-3 GeV  (which would exclude significant jet
quenching), to consistency with unity ~\cite{denterria}. The
latter situation might correspond to a trade--off between initial
state multiple scattering (Cronin effect) and final state high
$p_T$ suppression.

\begin{figure}[htb]
 \includegraphics[width=\linewidth]{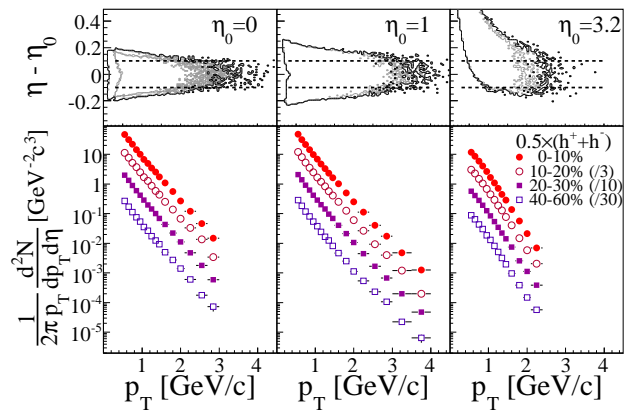}
  \caption{Spectra of charged hadrons measured with BRAHMS for the
  reaction $Au+Au$ at \snn{62.4} for the four indicated centrality
  classes at $\eta=0$ (left panels), $\eta=1$ (middle panels) and
  $\eta=3.2$ (right panels). Upper row: the distributions show the
  spectrometer acceptance at the various settings.
  The dotted lines show the cuts in $\eta$ that have been applied.}
  \label{fig:spec}
\end{figure}

The data presented here have been collected with the BRAHMS
detector at RHIC~\cite{brahms_nim}. BRAHMS consists of a detector
system for triggering and for vertex and centrality determination
and two magnetic spectrometers for precision measurements of
charged hadrons. For the measurements presented here the
midrapidity spectrometer was positioned at 90 and 45 degrees
relative to the beam direction corresponding to average
pseudorapidity $\eta=0$ and $\eta=1$ and the forward spectrometer
was positioned at 3 degrees enabling measurements at $\eta=3.2$
(see actual acceptances in the top panels of fig. \ref{fig:spec}).

The trigger permitted the recording of events corresponding to
approximately 90\% of all inelastic $Au+Au$ collisions at
\snn{62.4}. The collision centrality has been determined from the
measured charged particle multiplicity in a detector covering
$|\eta|<2$. The collision vertex is determined from the difference
in arrival time of particles scattered at small angles into two
arrays of Cherenkov radiator counters situated on either side of
the nominal interaction point around the beam pipe. The momentum
of charged hadrons is obtained by determining their trajectories
through the magnetic spectrometers using time projection chambers
and position sensitive drift chambers.

Figure \ref{fig:spec} shows the measured spectra for four
different centrality classes at $\eta=0$, $\eta=1$ and $\eta=3.2$.
The upper row shows the corresponding acceptance of the
spectrometers at the three nominal angular settings (90, 45 and 3
degrees). The spectra have been obtained by counting the number of
measured charged particle tracks per event in the spectrometers in
each of the centrality ranges and have been corrected for the
finite acceptance of the spectrometers, the efficiency of the
tracking chambers and for smearing due to the momentum resolution.
Corrections to spectra for feed--down from weakly decaying
particles have not been applied since the yields of different
species of particles are not directly measured. The systematic
errors on the normalization of the spectra is estimated to be
$\approx 10\%$. In general, the analysis is similar to that
described in ref.~\cite{brahms_highpt}.

\begin{figure*}[htb!]
  \includegraphics[width=0.99\linewidth]{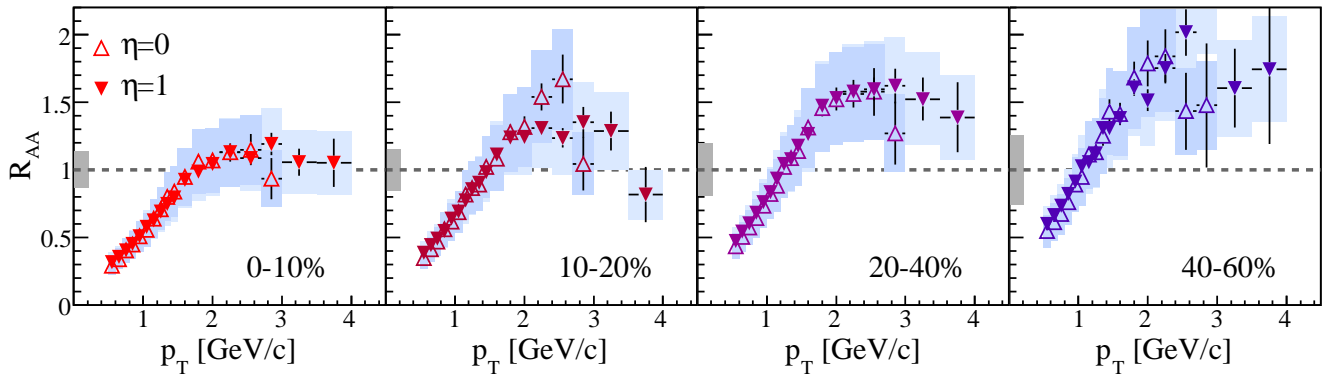}
  \caption{Nuclear modification factor at $\eta=0$ (open triangles)
    and $\eta=1$ (closed triangles) for the four centrality classes
    indicated. The magnitude of the uncertainty on $\langle N_{bin} \rangle$ is
    indicated with the shaded band around unity. The estimated systematic
    errors on the data points (from normalisations and reference spectra) are
    indicated by the shaded boxes around the points.
    The $\pm 20\%$ uncertainty on the p+p reference spectrum affects
    all the shown ratios equally.}
  \label{fig:raa}
\end{figure*}

In Figure~\ref{fig:raa} we display the nuclear modification
factors $R_{AuAu}$ as function of $p_T$ for the different
centrality classes at $\eta=0$ and $\eta=1$. The nuclear
modification factor is defined as:
$R_{AuAu}=\frac{{d^2N^{AuAu}/dp_Td\eta}}{\langle N_{bin}\rangle
d^2N^{NN}/dp_Td\eta}$. It involves the scaling of the $p_T$
spectra from elementary nucleon-nucleon (in practice
proton--proton) collisions by the mean number of binary
nucleon--nucleon collisions, $\langle N_{bin} \rangle$. For the
different centrality classes $\langle N_{bin} \rangle$ has been
estimated using HIJING events including the constraints imposed by
the actual detector geometry and resolution. We have used $\langle
N_{bin} \rangle$=752$\pm100$, 459$\pm70$, 217$\pm42$ and 70$\pm18$
for the centrality classes $0-10\%$, $10-20\%$, $20-40\%$ and
$40-60\%$, respectively. The uncertainties arise both from the
estimated uncertainty on the efficiency of our min bias trigger
($\pm 5\%$) and from estimates of the dependence of $\langle
N_{bin} \rangle$ on variations of the parameters for the Glauber
model calculation.

At \snn{62.4}, $p+p$ collisions have not yet been measured at
RHIC. Thus, in order to construct the $R_{AuAu}$ we must rely on
results from earlier experiments. In particular, we have based the
analysis on parametrizations of $p+p$ data~\cite{ISRdata} obtained
at the ISR collider at \snn{53} and \snn{62.4}. The ISR
distributions have been scaled to the present energy and rapidity
range using scaling factors obtained by simulating events using
the HIJING $p+p$ event generator. In the present work, we have
parameterized the measured yield by a power law $Ed^3\sigma/d^3p =
A \times (1 + p_T/p_0)^{-n}$. For $\eta=0$ we use
$A=\unit[244.2]{mbGeV^{-2}c^3}$, $p_0=\unit[2.00]{GeV/c}$ and
$n=14.31$, for $\eta=1$ we use $A=\unit[222.6]{mbGeV^{-2}c^3}$,
$p_0=\unit[2.16]{GeV/c}$ and $n=15.07$. The inelastic cross
section is set to $\sigma=\unit[36]{mb}$. A more detailed study of
the reference spectrum at \snn{62.4} can be found in
ref.~\cite{denterria_ref}. Our parametrization is consistent with
the results quoted therein. We assign a $\pm$20\% systematic
uncertainty to the reference spectrum.

Figure~\ref{fig:raa} shows that the $R_{AuAu}$ for $\eta=0$ and
$\eta=1$ are similar within each of the 4 considered centrality
classes. For the more peripheral collisions a pronounced
enhancement above unity of the nuclear modification is seen for
$p_T > \unit[1] {GeV/c}$. It is much larger than the corresponding
enhancement of $R_{AuAu}$ seen for $Au+Au$ collisions at \snn{200}
for the same centrality class ($40-60\%$). As the collisions
become more central, the $R_{AuAu}$ decreases systematically. For
the $0-10\%$ centrality bin the measurements are consistent with
binary scaling at $p_T=\unit[4]{GeV/c}$. Even considering the
large systematic errors associated with the reference spectra and
the normalization to $\langle N_{bin} \rangle$ the $R_{AuAu}$
around midrapidity at \snn{62.4} is significantly above the value
obtained for $Au+Au$ collisions at \snn{200} (where $R_{AuAu}
\approx 0.4 $ at $p_T = \unit[4]{GeV/c}$).

\begin{figure}[htb]
  \includegraphics[width=0.99\columnwidth]{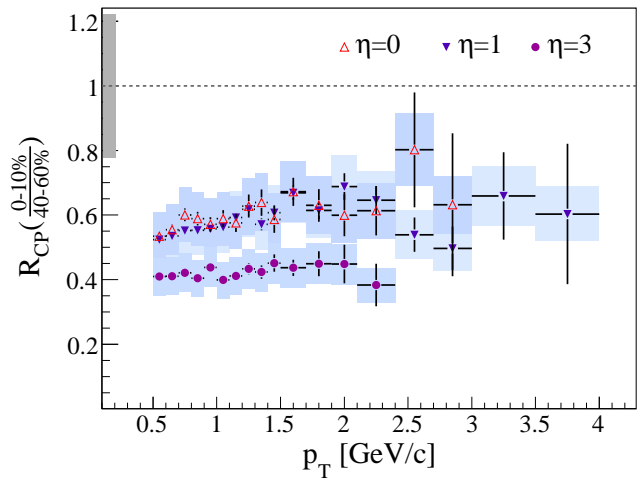}
  \vspace{-0.3cm}
  \caption{$R_{CP}$ ratio for centrality ratios (0-10\% to 40-60\%)
  for the three considered rapidities: $\eta=0$ (open triangles),
  $\eta=1$ (closed triangles), $\eta=3.2$ (closed circles). The
  systematic errors are indicated by the shaded boxes.}
  \label{fig:rcp}
\end{figure}

In order to avoid the dependence on the $p+p$ reference we
construct the $R_{CP}$ ratio of the $R_{AuAu}$ for two
centralities, in the present case \ $0-10\%$ and $40-60\%$.The
distributions of $R_{CP}$ around midrapidity ($\eta=0$ and
$\eta=1$) are shown in figure \ref{fig:rcp}. They exhibit similar
values (around 0.6), and are similar to the ratios that have been
measured for \snn{200} collisions. $R_{CP}$ ratios have also
recently been studied at SPS by the NA57 collaboration at central
rapidity ~\cite{NA57-RCP}. In that work, the $R_{CP}$ is above
unity for $p_T > 2 $GeV/c for negative hadrons. The ratios
measured in this work decrease, however, to $R_{CP} \approx 0.4$
as the pseudorapidity increases to $\eta=3.2$. At this high
pseudorapidity, the $R_{CP}$ shows little $p_T$ dependence and the
level of the $R_{CP}$ is similar to that observed at the same
pseudorapidity for $d+Au$ collisions~\cite{brahms_rda_vs_eta}. We
note that for the $\eta =3.2$ setting, the average angle of
observation is about 4.5 degrees. At this angle the transverse
momentum corresponding to scattering of a particle with full beam
momentum is $p_T \approx  \unit[62.4]/2 \cdot sin(4.5)$ {GeV/c} $=
\unit[2.44]{GeV/c}$. Thus, even though the absolute $p_T$ scale is
moderate, particles in this $p_T$ range at forward rapidities have
high momentum. It is remarkable that there is yield practically up
to this kinematic limit, suggesting that the yield at the high end
of the spectrum, at the most forward angles, may be dominated by
nearly inelastically scattered particles. We have also studied the
$R_{CP}$ ratios separately for positive and negative hadrons, but
no significant differences were found.

\begin{figure}[htb]
  \includegraphics[width=0.99\columnwidth]{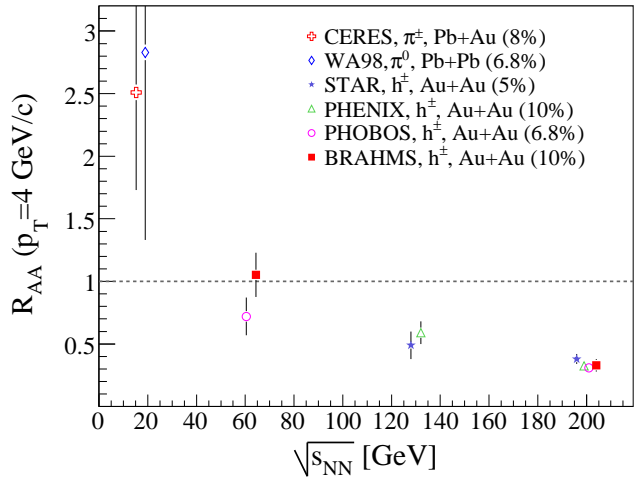}
  \vspace{-0.1cm}
  \caption{Energy systematics of the nuclear modification factor in
    central heavy ion collisions at $p_T=\unit[4]{GeV/c}$ for
    \snn{17.2,62.4,130 \text{ and } 200} at midrapidity (data from the same energy
    range have been offset for clarity). The data is from
    ref.~\cite{wa98_raa,ceres_raa,star_rauau_200,phenix_rauau_200,phobos_rauau_200,star_rauau_130,phenix_rauau_130,brahms_highpt}.}
  \label{fig:esyst}
\end{figure}

In figure \ref{fig:esyst} we show the systematics of the nuclear
modification factor around midrapidity for central
nucleus--nucleus collisions at $p_T=\unit[4]{GeV/c}$ as a function
of beam energy. The values for the SPS energy range (which are for
pions produced in central for Pb+Pb collisions) are assigned a
large error reflecting the poorly known $p+p$ reference also at
that energy. The figure indicates that the $R_{AuAu}$ for
\snn{62.4} is higher than for $Au+Au$ reactions at higher energy,
but below the values from the SPS, suggesting a transition in the
degree of high $p_T$ suppression from the low energy regime to the
high energy regime. It is, however, not possible to conclude (
primarily due to the important uncertainties in the $p+p$
reference spectra for \snn{62.4} and \snn{17.2}) whether this
transition occurs gradually from SPS energies or has the character
of a more sudden onset between SPS and RHIC(\snn{62.4}).

The present data raise a number of important issues pertaining to
the understanding of high $p_T$ suppression in heavy ion
collisions. At the experimental level the discussion presented
here highlights the need for accumulating reliable $p+p$ reference
data at RHIC for all ion-ion energies studied. The present data
show that suppression diminishes with decreasing energy at
midrapidity, an observation that calls for detailed theoretical
study of the energy dependence of jet suppression in the parton
energy loss model. At the same time we find that high $p_T$
suppression is not confined to midrapidity but appears to be a
general phenomenon over a large rapidity and energy range. It may
be, that the significant high $p_T$ suppression which is seen at
forward angles (high rapidity) reflects a longitudinally extended
suppressing medium -- although this has yet to be validated by
theoretical investigations (see e.g.~\cite{HiranoNara},
\cite{Barnafoldi05}). It may also be that a second suppressing
mechanism is at play at forward rapidities. The Color Glass
Condensate model predicts one such mechanism related to low--x
gluon saturation. In any case a consistent understanding of the
high $p_T$ phenomenon across a broad range of rapidity and energy
is needed.

This work was supported by the Office of Nuclear Physics of the
U.S. Department of Energy, the Danish Natural Science Research
Council, the Research Council of Norway, the Polish State
Committee for Scientific Research (KBN) and the Romanian Ministry
of Research.

\end{document}